\documentclass[sigconf,nonacm,dvipsnames]{acmart}  
\usepackage{multirow}
\usepackage{subcaption}

\usepackage{tcolorbox}
\usepackage{newtxtext,newtxmath}
\usepackage{enumitem}

\usepackage{pifont}
\usepackage{flushend}

\setcopyright{none}                      

\begin{document}
\title{Reinforcement Fine-Tuning for Reasoning towards Multi-Step Multi-Source Search in Large Language Models}

\author{Wentao Shi}
\email{wentao.shi0429@gmail.com}
\affiliation{%
  \institution{Toursun Synbio}
  \city{Shanghai}
  \country{China}
}

\author{Yiqing Shen}
\authornote{Corresponding author.}                     
\email{yshen92@jhu.edu}
\affiliation{%
  \institution{Johns Hopkins University}
  \city{Baltimore}
  \state{Maryland}
  \country{USA}
}

\begin{abstract}
Large language models (LLMs) can face factual limitations when responding to time-sensitive queries about recent events that arise after their knowledge thresholds in the training corpus. 
Existing search-augmented approaches fall into two categories, each with distinct limitations: multi-agent search frameworks incur substantial computational overhead by separating search planning and response synthesis across multiple LLMs, while single-LLM tool-calling methods restrict themselves to sequential planned, single-query searches from sole search sources.
We present Reasoning-Search (R-Search), a single-LLM search framework that unifies multi-step planning, multi-source search execution, and answer synthesis within one coherent inference process.
Innovatively, it structure the output into four explicitly defined components, including reasoning steps that guide the search process (\texttt{<think>}), a natural-language directed acyclic graph that represents the search plans with respect to diverse sources (\texttt{<search>}), retrieved results from executing the search plans (\texttt{<result>}), and synthesized final answers (\texttt{<answer>}). 
To enable effective generation of these structured outputs, we propose a specialized Reinforcement Fine-Tuning (ReFT) method based on GRPO, together with a multi-component reward function that optimizes LLM's answer correctness, structural validity of the generated DAG, and adherence to the defined output format. 
Experimental evaluation on FinSearchBench-24, SearchExpertBench-25, and seven Q\&A benchmarks demonstrates that R-Search outperforms state-of-the-art methods, while achieving substantial efficiency gains through 70\% reduction in context token usage and approximately 50\% decrease in execution latency. 
Code is available at \url{https://github.com/wentao0429/Reasoning-search}.
\end{abstract}

\maketitle

\section{Introduction}\label{sec:intro}

Although large language models (LLMs) demonstrate zero-shot generalization in various tasks, they still face temporal and factual limitations when addressing questions about recent events, evolving facts, and time-sensitive developments that arise after their knowledge thresholds~\cite{lewis2020retrieval,gao2023retrieval,zhao2024retrieval,vu-etal-2024-freshllms}. 
These limitations stem from the static nature of their training corpus, which becomes increasingly outdated as real-world information evolves. 
Therefore, LLMs increasingly require complex search capabilities to bridge this knowledge gap, enabling them to access current information, verify facts against authoritative sources, and provide accurate responses that reflect the latest developments~\cite{xiong2024searchengineservicesmeet,ma2024craftingknowledgeexploringcreative}.
Previous research has looked into two directions to enable searching with LLMs, namely (1) multi-agent approaches that involve interactions across multiple LLMs and (2) tool-calling methods with a single LLM.

Multi-agent search frameworks~\cite{chen2024mindsearchmimickinghumanminds,li2024agentframeworkrealtimefinancial,searchexpert,yang2025demystifyingenhancingefficiencylarge,jiang2024mmsearch} typically employ a three-stage design: a planning stage that decomposes complex queries into structured search plans represented by DAGs, an execution stage that implements these plans through API calls to external search engines, and a synthesis stage that aggregates retrieved information into coherent responses. 
Representative methods include \textsc{MindSearch}~\cite{chen2024mindsearchmimickinghumanminds} that pioneered this direction through graph-based search planning with parallel node execution, \textsc{FinSearch}~\cite{li2024agentframeworkrealtimefinancial} that specialized this approach for financial domains with temporal reasoning, \textsc{SearchExpert}~\cite{searchexpert} which introduced hierarchical planning for complex multi-hop queries, and \textsc{SearchAgent-X} enhanced efficiency through adaptive search depth control~\cite{yang2025demystifyingenhancingefficiencylarge}. 
However, this three-stage design in these multi-agent search frameworks inherently introduces limitations that undermine their effectiveness in real-world search applications. 
First, the separation of planning, execution, and synthesis stages across multiple LLMs dramatically increases computational overhead, as each stage requires separate LLM invocations with full context passing, resulting in token consumption that scales multiplicatively with search complexity. 
This becomes particularly problematic for time-sensitive applications where users expect rapid responses, as the cumulative latency from sequential LLMs calls can exceed acceptable thresholds for interactive search experiences. 
Second, delegation of the final generation of answers to a separate synthesis creates a ``\textit{plan-execution gap}'' where the synthesizing LLM lacks direct access to the original planning rationale, leading to responses that may contradict or incompletely reflect the executed search strategy.
Finally, these agents suffer from ``\textit{planning without reasoning}'', where the planning agent generates search plans without reasoning, while the reasoning module synthesizes results without knowing the search logic.

In contrast, tool-calling search methods with a single LLM adopt a different approach by integrating search capabilities directly within the LLM reasoning process. 
For example, \textsc{Search-R1}~\cite{jin2025search} extends reinforcement learning (RL) to train LLMs that generate multiple search keywords during the step-by-step reasoning of LLM. 
Similarly, \textsc{R1-Searcher}~\cite{song2025r1searcherincentivizingsearchcapability} employs a two-stage outcome-based RL approach that enables LLMs to invoke an external search without requiring process rewards or human-annotated data. 
\textsc{ZeroSearch}~\cite{sun2025zerosearch} addresses the search API cost constraints using simulated search during training, where an auxiliary LLM generates relevant and noisy documents to train the policy model without real search engine interactions. 
Although these approaches offer computational efficiency for agent-based methods and straightforward implementation, they encounter additional constraints when addressing complex, multi-faceted search requirements.
First, their training and inference typically depend on offline knowledge bases or simulated environments rather than live search engines, severely constraining their ability to access genuine real-time information. 
Additionally, these methods operate through sequential, single-query search that process one query at a time without leveraging inter-query dependencies, which therefore limits their ability to pursue complex information searching where multi-query-based searching is necessary.
Most critically, these approaches remain confined to a homogeneous information source, typically a single search API or knowledge base, preventing them from leveraging the complementary strengths of diverse information repositories such as specialized search sources, like academic archives, real-time news feeds, and domain-specific search engines that human experts naturally consult when conducting a comprehensive search.

The major limitation of existing approaches lies in their partial solutions, where multi-agent methods achieve global planning at the cost of computational complexity and architectural brittleness, while tool-calling with a single LLM maintains efficiency but sacrifices strategic depth and multi-source search capabilities. 
To address these limitation, we present \textbf{Reasoning-Search (R-Search)}, a novel single-LLM search framework that unifies multi-step planning, multi-source search execution, and answer synthesis within one coherent inference process. 
R-Search addresses the limitations of both paradigms by structuring its output into four explicitly defined components: reasoning steps that guide the search process (\texttt{<think>}), a natural-language directed acyclic graph (NL-DAG) that encodes multi-node search plans across diverse online sources (\texttt{<search>}), retrieved results from executing the search plans (\texttt{<result>}), and synthesized final answers (\texttt{<answer>}). 
To train LLMs that effectively generate these structured outputs, we propose a \textbf{Reinforcement Fine-Tuning (ReFT)} specific for LLM searching based on the GRPO \cite{deepseekai2025deepseekr1incentivizingreasoningcapability}. 
Correspondingly, we introduce a multi-component reward function that simultaneously optimizes answer correctness, structural validity of the generated NL-DAG, and adherence to the defined output format to ensure that the LLM not only produces accurate answers but also generates executable search plans that can be reliably parsed and validated. 
%

The major contributions are three-fold.
First, we present R-Search, which unifies reasoning, planning, and synthesis in a single LLM inference process through structured output: reasoning traces (\texttt{<think>}), search plan DAGs (\texttt{<search>}), search execution results (\texttt{<result>}), and answers (\texttt{<answer>}). 
This eliminates architectural fragmentation while supporting multi-source searching. 
Second, we propose a specific ReFT for searching based on GRPO with a reward function that jointly optimizes the accuracy of the answer, the executability of the DAG, and the compliance of the format, ensuring that LLM generates both correct responses and reliable search plans. 
Third, we design an automated data construction method for the proposed ReFT to enable a fully automated training process.

\section{Related Work}\label{sec:related}

\subsection{Agent-Based Search Methods}
Multi-agent search frameworks typically employ coordinated LLMs to collectively decompose complex queries, execute searches, and synthesize final responses. 
MindSearch~\cite{chen2024mindsearchmimickinghumanminds} made the first attempt by decomposing user queries into smaller, manageable sub-queries represented by directed acyclic graph (DAG), where nodes represent independent web searches for specific sub-queries and edges denote reasoning relationships and topological dependencies between these searches. 
Similarly, FinSearch~\cite{li2024agentframeworkrealtimefinancial} extends MindSearch to financial domains by incorporating a finance-specific search tools and a temporal weighting mechanism that prioritizes the relevance of information based on the context of time.
SearchExpert~\cite{searchexpert} focus on the planning stage through training LLM through a combination of supervised fine-tuning (SFT) for initial policy learning and reinforcement learning (RL) for iterative optimization.
SearchAgent-X~\cite{yang2025demystifyingenhancingefficiencylarge} introduces adaptive search depth control to optimize the balance between search comprehensiveness and computational efficiency, which can perform shallow searches for simple factual queries while allocating deeper, more exhaustive searches for complex multi-hop questions that require extensive information gathering.
However, all of these agent-based search methods suffer from high computational overhead due to separation of planning, search execution, and answer synthesis across multiple LLMs.
Additionally, the multi-stage design creates the plan-execution gap as it limits their ability to adapt dynamically to intermediate search results, leading to responses that may deviate from the original search strategy.
Finally, these agents suffer from the limitation of planning without reasoning, confining their performance to complex search queries.
These limitations of agent-based search methods, including computational inefficiency and architectural fragmentation, motivate the need for a unified framework that can preserve the sophistication of planning while integrating search and reasoning within a single model, which is the core design principle behind our R-Search approach.

\subsection{Single-LLM Search Methods}
In contrast to multi-agent search methods, single-LLM search methods embed search capabilities directly within the reasoning process of LLMs.
Search-R1~\cite{jin2025search} employs RL to train LLMs that autonomously generate search queries in a sequential manner during step-by-step reasoning, optimizing search-reasoning trajectories.
R1-Searcher~\cite{song2025r1searcherincentivizingsearchcapability} advances this approach with a two-stage outcome-based RL that enables LLMs to invoke external search engines without requiring process rewards or human-annotated data.
ZeroSearch~\cite{sun2025zerosearch} addresses the prohibitive costs of API-based search training by introducing a simulated search environment, where an auxiliary LLM generates both relevant and noisy documents to train the policy model.
However, all of these single-LLM methods remain limited to sequential, single-query search patterns that process one query at a time without leveraging inter-query dependencies or parallel search strategies.
Furthermore, their training and evaluation are primarily based on offline libraries or minimal online evaluation, severely limiting their response to recent events and time-sensitive information.
Finally, all these methods remain confined to sole search sources, typically a single search API or knowledge base, and lack explicit planning structures, impairing their interpretability and ability to perform sophisticated multi-source information gathering.
These limitations of single-LLM methods highlight the need for a framework that can maintain computational efficiency while enabling explicit multi-step planning and multi-source search capabilities, which R-Search achieves through its structured output format.

\section{Methods}\label{sec:method}

\subsection{Method Overview}
In this section, we elaborate on the Reasoning-Search (R-Search), a unified single-LLM search framework to enable integrated reasoning and multi-step multi-source search execution without the architectural fragmentation and computational overhead inherent in existing approaches.
While multi-agent search methods~\cite{chen2024mindsearchmimickinghumanminds,li2024agentframeworkrealtimefinancial,searchexpert,yang2025demystifyingenhancingefficiencylarge} achieve planning through architectural fragmentation across multiple LLMs, and single-LLM search approaches~\cite{jin2025search,song2025r1searcherincentivizingsearchcapability,sun2025zerosearch} maintain efficiency at the cost of strategic depth, R-Search unifies multi-step reasoning, structured search planning, and answer synthesis within a single coherent inference process by one LLM.
Central to R-Search is a structured output format comprising four explicitly defined components, including reasoning traces (\texttt{<think>}) to clarify information needs, search plans in terms of natural-language directed acyclic graphs (NL-DAG) (\texttt{<search>}), retrieved results (\texttt{<result>}), and final answers (\texttt{<answer>}). 
This structured representation eliminates the need for separate planning and synthesis modules while supporting parallel search execution across heterogeneous information sources.
To adapt LLMs to this structured search paradigm, we introduce ReFT based on GRPO. 
Our approach employs a multi-component reward function that jointly optimizes answer accuracy, structural validity of the generated NL-DAG, and adherence to the defined output format. 
Additionally, we propose an automated dataset construction method for ReFT that generates multi-hop questions requiring reasoning and multi-source searching, ensuring the model develops capabilities beyond simple keyword matching or single-source retrieval.

\subsection{Task Formulation}

Given a user query $q \in \mathcal{Q}$ and a set of available search tools $\mathcal{T}$ (\textit{e.g.}, news search, academic paper search, general web search), our objective is to train a LLM $\pi_\theta$ that performs integrated reasoning, search planning, and final answer synthesis with one inference pass. 
To be more specific, the LLM must jointly perform four interconnected tasks: (1) reasoning to analyze the query and identify required information, (2) planning to construct a multi-source search strategy, represented by an NL-DAG, (3) execution to interface with external search tools through environment $\mathcal{E}$, (4) and synthesis to generate answers from retrieved evidence.
Therefore, the LLM is trained to produces a structured output $y = (r, G, R, a)$ given $q$ by following a fix template:
\begin{equation}
\begin{aligned}
y = \; & \underbrace{\texttt{<think>} \; r \; \texttt{</think>}}_{\text{reasoning}}
\;
\underbrace{\texttt{<search>} \; G \; \texttt{</search>}}_{\text{planning}} \\
& \underbrace{\texttt{<result>} \; R \; \texttt{</result>}}_{\text{execution}}
\;
\underbrace{\texttt{<answer>} \; a \; \texttt{</answer>}}_{\text{synthesis}}
\end{aligned}
\label{eq:format}
\end{equation}
where $r$ represents free-form reasoning, $G = (V, E)$ denotes the search plan in terms of NL-DAG with nodes $V$ as query-tool pairs and edges $E$ as dependencies, $R$ contains concatenated search results from execution $G$, and $a$ is the synthesized final answer. 
Structured tags $\texttt{<tag>}$ and $\texttt{</tag>}$ are learnable tokens that the LLM generates to demarcate different components.
During inference, $\pi_\theta$ generates tokens autoregressively until producing token \texttt{</search>}. 
The environment then parses $G$, executes searches in topological order with parallelization of search API calls where possible, and appends the execution results between \texttt{<result>} and \texttt{</result>} before resuming generation until \texttt{</answer>}.
To optimize LLM to generate this structured output for searching, our training objective maximizes the expected reward over a dataset $\mathcal{D} = \{(q_i, a^i)\}_{i=1}^{|\mathcal{D}|}$ of query-answer pairs, where each $q_i$ represents a user query involving search and $a^i$ denotes the corresponding ground-truth answer. 
The optimization seeks parameters $\theta^*$ that maximize expected performance across all possible query-answer-generation triplets:
\begin{equation}
\theta^* = \arg\max_{\theta} \mathbb{E}_{(q, a) \sim \mathcal{D}, y \sim \pi_\theta(\cdot|q)} 
\left[ \mathcal{R}(y, a) \right]
\end{equation}
where $\pi_\theta(\cdot|q)$ represents the probability distribution over possible structured outputs $y$ given query $q$ under the current LLM $\pi$ parameterized by $\theta$, and $\mathcal{R}(y, a)$ denotes the reward function that evaluates the quality of generated output $y$ against ground-truth answer $a$.
The composite reward function $\mathcal{R}(y, a)$ decomposes the evaluation into three orthogonal dimensions, depicted as:
\begin{equation}
\mathcal{R}(y, a) = 
\alpha_{\text{fmt}} \cdot \mathcal{F}_{\text{fmt}}(y) 
+ \alpha_{\text{dag}} \cdot \mathcal{F}_{\text{dag}}(G) 
+ \alpha_{\text{ans}} \cdot \mathcal{F}_{\text{ans}}(y,a),
\label{eq:objective}
\end{equation}
where $\alpha_{\text{fmt}}, \alpha_{\text{dag}}, \alpha_{\text{ans}} \in [0,1]$ with $\alpha_{\text{fmt}} + \alpha_{\text{dag}} + \alpha_{\text{ans}} = 1$ represent weighting coefficients that balance the importance of format compliance ($\mathcal{F}_{\text{fmt}}$), NL-DAG structural validity ($\mathcal{F}_{\text{dag}}$), and answer accuracy ($\mathcal{F}_{\text{ans}}$) respectively. 
This decomposition ensures that the LLM learns not only to generate correct final answers, but also to produce well-structured, executable search plans that can be reliably parsed and executed by the environment.

\subsection{Natural-Language DAG Representation and Execution}

\begin{figure}[htbp]
  \centering
  \includegraphics[width=\linewidth]{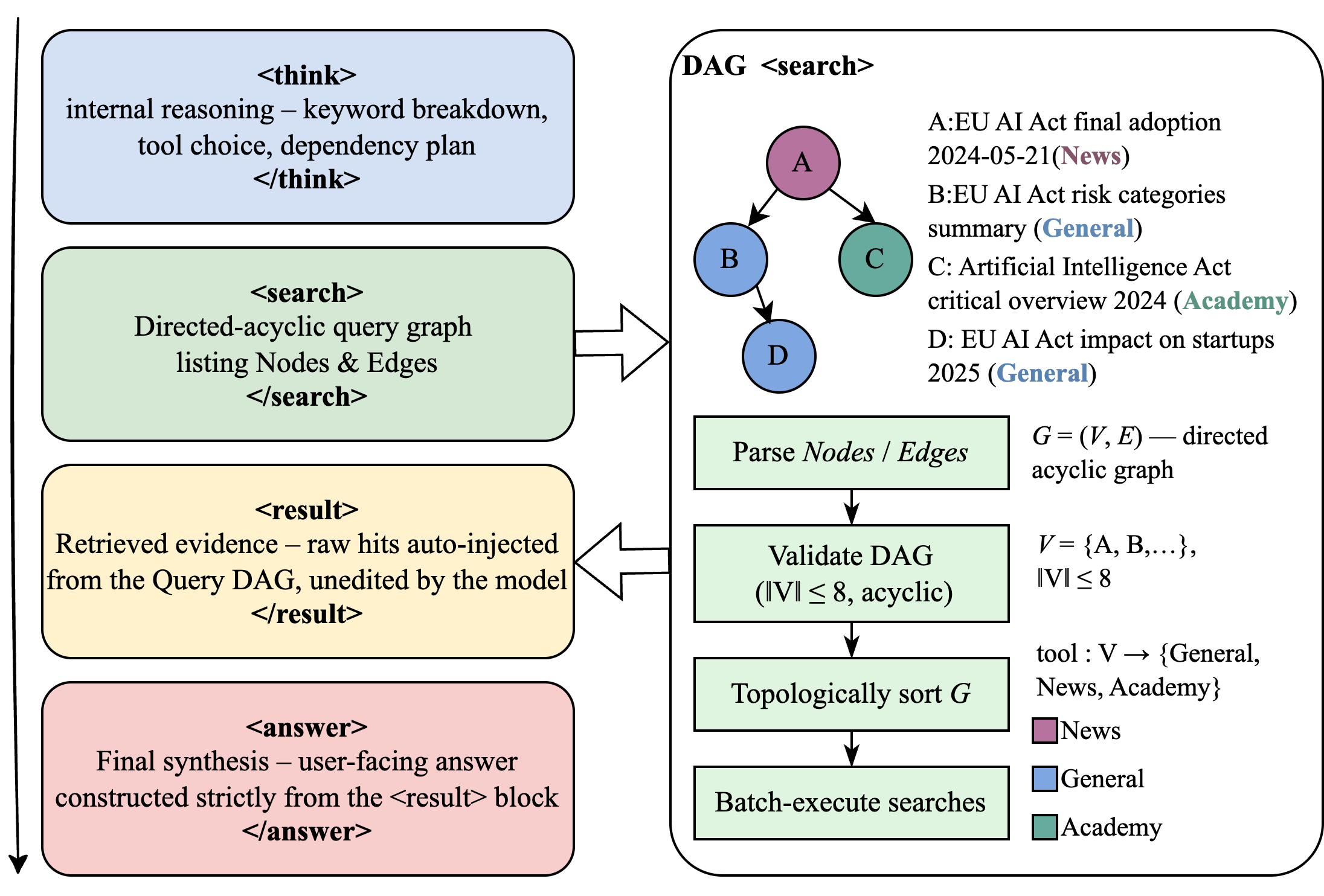}
  \caption{Illustration of the R-Search structured output format and execution pipeline. 
  The left panel illustrates the four-component template where the LLM generates reasoning traces in \texttt{<think>}, constructs an NL-DAG in \texttt{<search>}, receives retrieved evidence in \texttt{<result>}, and produces synthesized answers in \texttt{<answer>}. 
  The right panel demonstrates the automated NL-DAG execution process, including node-edge parsing, structural validation with acyclicity constraints, topological sorting, and parallel search execution across heterogeneous information sources.}
  \label{fig:searchTEMP}
\end{figure}

One major innovation of R-Search lies in its representation of a natural-language directed acyclic graph (NL-DAG) for the search plan within the structured output format $y$ defined in Eq.~\eqref{eq:format}. 
As illustrated in Fig.~\ref{fig:searchTEMP}, this representation enables the LLM to integrate search planning with reasoning while maintaining computational efficiency and interpretability.
Importantly, the NL-DAG representation offers advantages over traditional code-based or JSON formats in previous works \cite{chen2024mindsearchmimickinghumanminds,li2024agentframeworkrealtimefinancial}, as the NL-DAG typically requires only fewer tokens. 
The reasoning component $r$ in the \texttt{<think>} stage serves as the foundation for search planning, where the LLM explicitly analyzes the user query $q$ to identify information gaps, decompose complex queries into sub-queries, and establish the conceptual framework for subsequent search operations. 
This reasoning process directly informs the construction of the search plan $G = (V, E)$, ensuring that the generated NL-DAG addresses the specific information needs identified during analysis.

The NL-DAG structure $G$ within the \texttt{<search>} component employs a human-readable representation that maintains both semantic clarity and structural precision. 
Each node $v \in V$ is expressed as a pair of sub-query and search tool using natural language descriptors, such as ``\textit{A: EU AI Act final adoption 2024-05-21 (News)}'', where the identifier ``\textit{A}'' serves as a reference key for the node name, the sub-query ``\textit{EU AI Act final adoption 2024-05-21}'' specifies the information target need to be searched, and the parenthetical notation designates the search tool from the available set $\mathcal{T}$. 
The edge set $E$ in $G$ explicitly defines the dependencies between nodes through a concise syntax following the definitions of the nodes, utilizing semicolon-separated relationships such as ``\textit{A $\rightarrow$ B; C $\rightarrow$ B}'' to establish the topological structure.

The execution process for the generated NL-DAG $G$ follows an automated validation and execution pipeline. 
Upon completion of the \texttt{<search>} generation, the environment $\mathcal{E}$ employs regular expression parsing to extract nodes and edges, subsequently validating structural integrity to ensure acyclicity. 
It enforces practical constraints including maximum node limits and tool availability verification, where nodes referencing undefined search sources are flagged and excluded from execution. 
Following successful validation, a topological sort determines the execution order, enabling parallel dispatch of independent queries to their respective search endpoints within $\mathcal{T}$.
After the execution of the search plan, the search results are organized within the \texttt{<result>} component $R$, where passages are concatenated according to the execution order of the node and prefixed with the corresponding node identifiers. 
This structured alignment facilitates effective utilization of retrieved evidence during answer synthesis, as the LLM can easily associate search results with their originating queries and leverage dependency relationships established in the NL-DAG.

\subsection{Reinforcement Fine-Tuning for Search-Augmented Generation}

\begin{figure*}[htbp]
  \centering
  \includegraphics[width=\linewidth]{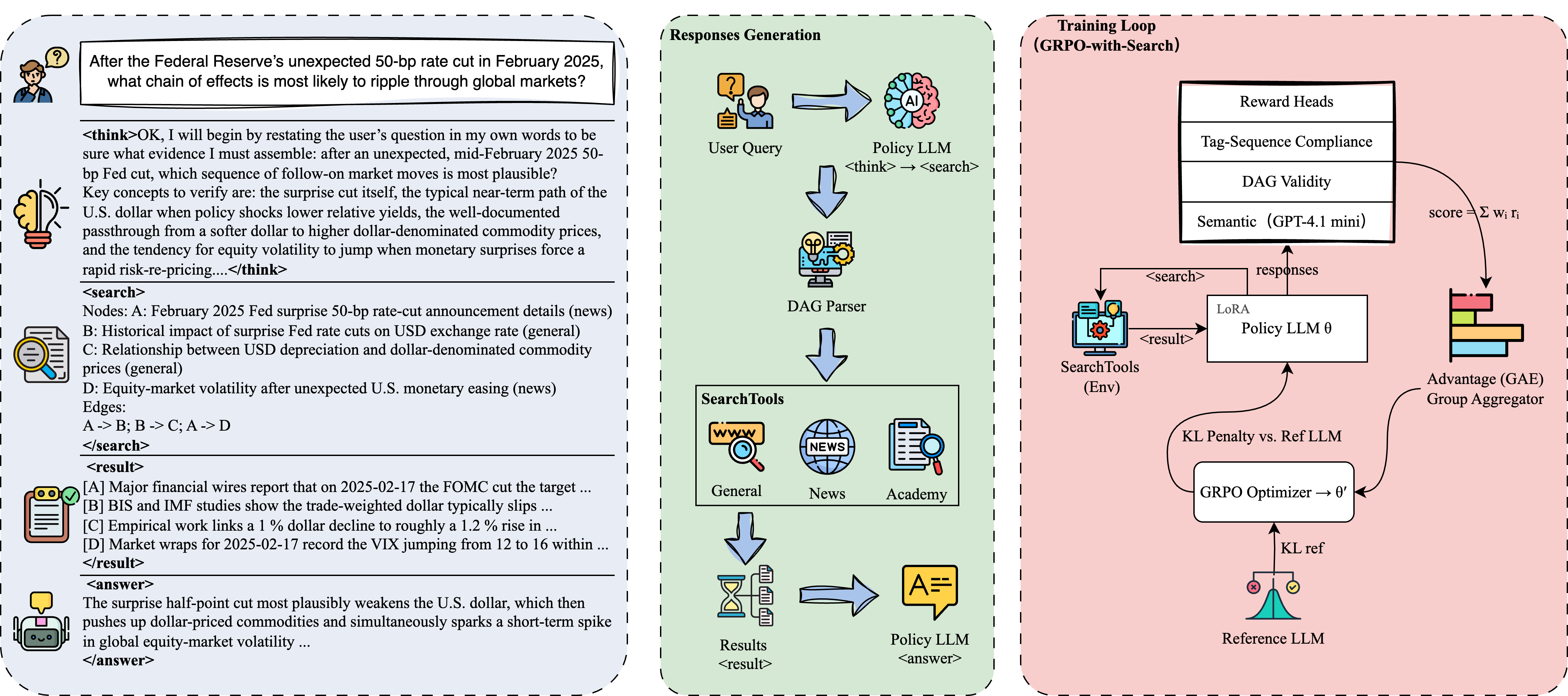}
  \caption{ReFT training pipeline illustrating the integration of structured generation with GRPO optimization. 
  The left panel demonstrates a complete inference example showing the four-component structured output generation (\texttt{<think>}, \texttt{<search>}, \texttt{<result>}, \texttt{<answer>}) for a complex financial query. 
  The center panel abstracts the response generation workflow, highlighting the interaction between the policy LLM, NL-DAG parser, and external search tools. 
  The right panel details the GRPO training loop, where the composite reward function combining format compliance, DAG validity, and answer accuracy drives iterative policy improvements through advantage estimation and KL-regularized optimization.}
  \label{fig:GRPOtrainer}
\end{figure*}

To optimize the LLM $\pi_\theta$ for generating the structured output format defined in Eq.~\eqref{eq:format}, we introduce a specialized ReFT approach based on GRPO~\cite{shao2024deepseekmathpushinglimitsmathematical,deepseekai2025deepseekr1incentivizingreasoningcapability}. 
This adaptation addresses the challenges of mixed discrete-environment interaction that arises when the LLM interfaces with external search tools during the structured generation process.
As illustrated in Fig.~\ref{fig:GRPOtrainer}, the proposed ReFT method integrates the structured generation process with GRPO optimization.

The training process follows the structured generation protocol in which the LLM produces tokens autoregressively in the prescribed order: 
\texttt{<think>} $\rightarrow$ \texttt{<search>} $\rightarrow$ \texttt{<result>} $\rightarrow$ \texttt{<answer>}. 
Generation is strategically paused after the \texttt{</search>} token to enable environment interaction, where $\mathcal{E}$ parses the generated NL-DAG $G = (V, E)$, validates its structural integrity, executes the specified search queries through appropriate search APIs, concatenates the retrieved passages into $R$ and resumes autoregressive generation until \texttt{</answer>}.
The optimization objective defined in Eq.~\eqref{eq:objective} is realized through a composite reward function that simultaneously evaluates three orthogonal dimensions of the structured output, where the weighting coefficients are set to $\alpha_{\text{fmt}} : \alpha_{\text{dag}} : \alpha_{\text{ans}} = 0.25 : 0.25 : 0.5$ to balance structural compliance with answer quality optimization.

The format compliance reward $\mathcal{F}_{\text{fmt}}(y)$ implements binary evaluation, returning 1 when all four required components appear exactly once in the correct sequential order, and 0 otherwise. 
This component ensures consistent adherence to the structured template, enabling reliable parsing and execution by the environment $\mathcal{E}$.
The DAG structural validity reward $\mathcal{F}_{\text{dag}}(G)$ evaluates the search plan generated across multiple validation criteria. 
The component verifies acyclicity to ensure executable topological ordering, validates node format compliance requiring each node to specify both query content and associated tool designation, and confirms tool availability by checking that all referenced search sources belong to the predefined set $\mathcal{T}$. 
The reward returns 1 only when all validation criteria are satisfied, encouraging the generation of well-formed and executable search plans; otherwise 0.
The answer accuracy reward $\mathcal{F}_{\text{ans}}(y,a)$ employs a reference judge model (also implemented by an LLM) following established principles in reinforcement learning from AI feedback (RLAIF)~\cite{ouyang2022traininglanguagemodelsfollow,lee2024rlaifvsrlhfscaling}.
For factual and reasoning-based queries, the judge LLM computes continuous similarity scores in the interval $[0,1]$ based on semantic alignment between the generated answer and ground-truth responses. 
For multiple choice questions, F1 scores are calculated between the predicted and correct option sets. 
This component ensures that the LLM learns an effective synthesis of accurate responses from the retrieved evidence.

GRPO groups $M$ independent rollouts sharing identical queries, utilizing the reference policy $\pi_{\text{ref}}$ as a baseline for the estimation of advantages. 
The advantage for trajectory $i$ is computed as:
\begin{equation}
A_i = \mathcal{R}(y_i, a_i) - \frac{1}{M}\sum_{j=1}^{M}\mathcal{R}(y_j, a_j),
\end{equation}
where $y_i$ represents the generated output for the trajectory $i$ and $a_i$ denotes the corresponding ground truth answer.
The training loss for the minibatch $\mathcal{B}$ combines the optimization of the policy gradient with the regularization of the KL divergence, depicted as
\begin{equation}
\mathcal{L}_{\text{GRPO}} = -\frac{1}{|\mathcal{B}|}\sum_{i\in\mathcal{B}} \left[ A_i \cdot \log\frac{\pi_\theta(y_i)}{\pi_{\text{ref}}(y_i)} - \beta \cdot \text{KL}(\pi_\theta(\cdot|q_i) \| \pi_{\text{ref}}(\cdot|q_i)) \right]
\end{equation}
where $\beta$ undergoes linear annealing throughout the training to balance exploration with optimization stability.
A critical aspect of the ReFT implementation involves selective gradient computation that focuses exclusively on tokens generated by the policy LLM. 
Specifically, gradients are computed only for tokens within the \texttt{<think>}, \texttt{<search>}, and \texttt{<answer>} components, while the externally retrieved content in the \texttt{<result>} block is explicitly masked from gradient calculations to ensure that loss computations and subsequent parameter updates reflect only decisions directly under the model's control, preventing the introduction of uncontrollable noise from external search engines that could lead to training instability or unintended optimization behaviors.

\subsection{Automated Dataset Construction for ReFT}

Effective training of R-Search through ReFT requires a specialized dataset that challenges the LLM to plan for complex multi-source search. 
Our automated dataset construction pipeline addresses three requirements: (1) queries must necessitate information synthesis from multiple heterogeneous sources to encourage complex NL-DAG generation, (2) queries should target recent developments beyond the LLM's knowledge cut-off to enforce active search engagement, and (3) the dataset should exhibit progressive complexity to support stable RL convergence.

\begin{figure*}[htbp]
  \centering
  \includegraphics[width=\textwidth]{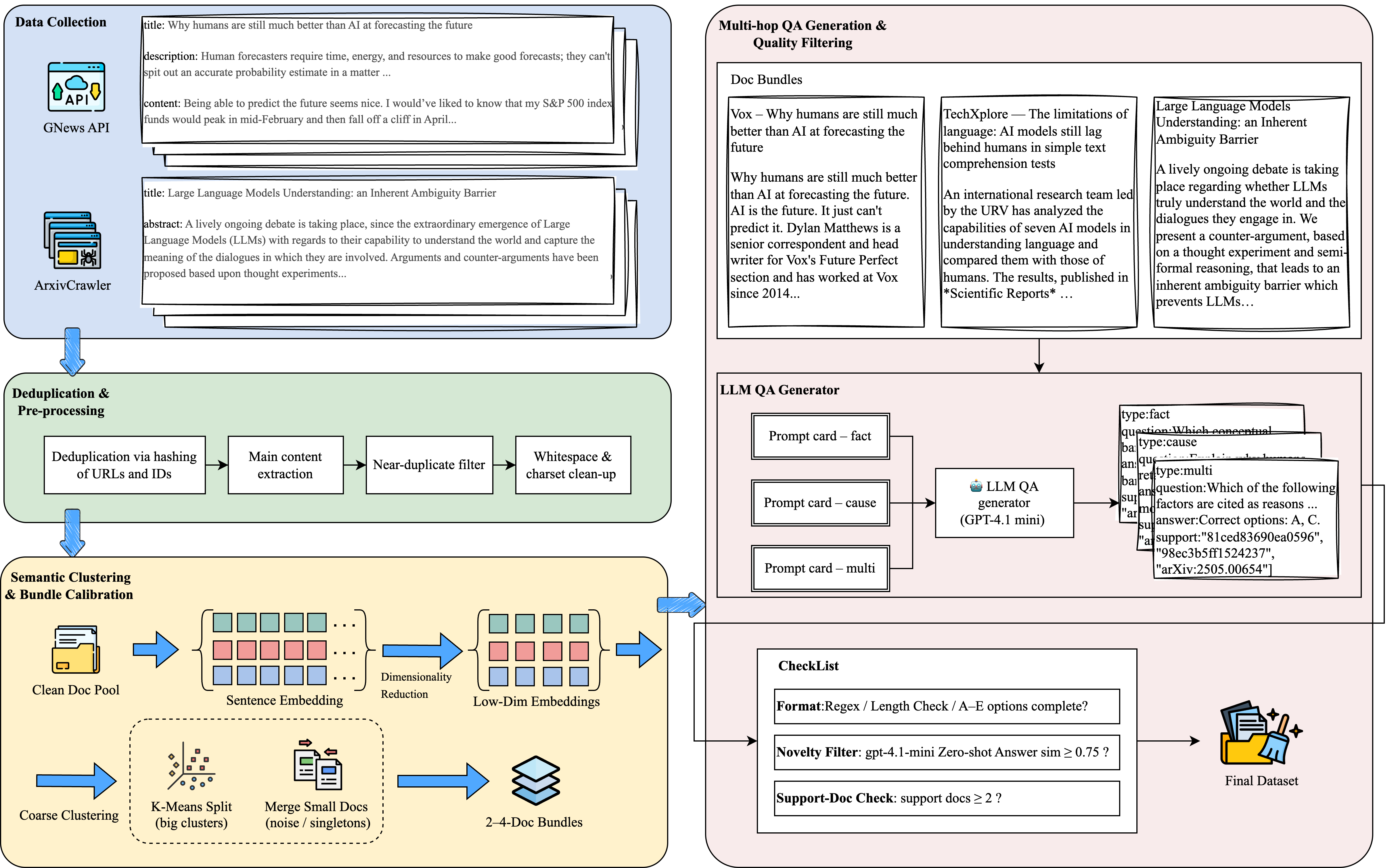}
  \caption{Automated multi-hop question-answering dataset construction pipeline for ReFT training. 
  The pipeline comprises four integrated stages: (1) data collection from diverse sources including news articles and academic papers, (2) deduplication and preprocessing to ensure content quality, (3) semantic clustering with dimensionality reduction to create thematically coherent document bundles, and (4) automated question-answer generation with quality filtering mechanisms that enforce multi-source reasoning requirements for ReFT training objectives.}
  \label{fig:pipeline:dataget}
\end{figure*}

As illustrated in Fig.~\ref{fig:pipeline:dataget}, our data collection strategy targets recent content from complementary domains within a three-month window from March 2025 to May 2025. 
The corpus comprises approximately 2,000 news articles that span current events, social developments, and policy updates, along with 1,500 recent arXiv manuscripts' abstracts from artificial intelligence and machine learning research domain. 
This combination ensures both topical diversity and temporal relevance, establishing conditions where successful question answering requires the multi-source search capabilities that R-Search is designed to optimize.
The collected documents undergo pre-processing to establish semantic coherence and eliminate redundancies. 
Following deduplication and content filtering, we employ semantic clustering to construct document bundles that enable realistic multi-hop reasoning scenarios \cite{lee2022deduplicatingtrainingdatamakes}. 
The documents are embedded using Sentence-BERT \cite{reimers-gurevych-2019-sentence} and dimensionally reduced through UMAP \cite{2018arXivUMAP} to 50 dimensions for computational efficiency. 
The initial agglomerative clustering organizes content into thematic groups, with oversized clusters recursively partitioned using K-means \cite{hartigan1979algorithm} until each bundle contains two to four semantically related yet source-diverse documents.
Multi-hop question-answer pairs are automatically generated from these document bundles using GPT-4.1-Mini \cite{openai2024gpt4technicalreport}. 
The generation process explicitly targets factual, causal, and multiple choice questions that inherently require evidence synthesis across multiple documents within each bundle. 
Critical filtering mechanisms ensure alignment with ReFT training objectives: (1) queries must not be answerable through pre-training knowledge alone, thereby compelling search tool utilization, and (2) each query must require genuine integration of evidence from multiple sources rather than simple keyword matching.
Following generation and filtering, 1,000 high-quality multi-hop question-answer pairs remain in the final dataset. 
To facilitate stable ReFT training, we organize these pairs according to complexity metrics, creating a curriculum that progresses from questions requiring fewer sources to more sophisticated multi-source reasoning scenarios. 
This prevents training instability during early ReFT stages while ensuring that the model develops robust search planning capabilities aligned with the composite reward function.

\section{Experiments}\label{sec:experiments}

\begin{table*}[t!]
\caption{Performance evaluation of R-Search against state-of-the-art search-augmented LLM method on FinSearchBench-24 and SearchExpertBench-25. 
It presents accuracy percentages, token consumption per answer, and end-to-end inference latency across baseline models without search augmentation, multi-agent search frameworks (SearchAgent, MindSearch, FinSearch), specialized search systems (SearchExpert), and commercial solutions (Perplexity Pro). 
Results represent mean $\pm$ standard deviation across multiple evaluation runs.
}
\label{tab:benchmarks}
\centering
\resizebox{\textwidth}{!}{%
\begin{tabular}{llcccccc}
\toprule
\multirow{2}{*}{\textbf{Method}}
& \multirow{2}{*}{\textbf{Models}}
& \multicolumn{3}{c}{\textbf{FinSearchBench-24\cite{li2024agentframeworkrealtimefinancial}}} 
& \multicolumn{3}{c}{\textbf{SearchExpertBench-25\cite{searchexpert}}} \\ 
\cmidrule(lr){3-5} \cmidrule(lr){6-8}
& 
& \textbf{Accuracy (\%)} 
& \textbf{Tokens (bit/ans)} 
& \textbf{Time (s/ans)}
& \textbf{Accuracy (\%)} 
& \textbf{Tokens (bit/ans)} 
& \textbf{Time (s/ans)} \\
\midrule
\multirow{5}{*}{Baseline}
& ChatGPT-4o \cite{yuan2024finllmsframeworkfinancialreasoning}        & 36.13 ± 1.22 & 293.82 ± 26.01 & \textbf{3.87 ± 0.15} & 22.50 ± 1.21 & 940.60 ± 45.98 & \textbf{6.04 ± 0.23} \\
& Llama-3.1-8B-instruct \cite{touvron2023llamaopenefficientfoundation} & 38.13 ± 1.27 & \textbf{210.44 ± 23.64} & 4.96 ± 0.25 & 21.00 ± 1.14 & \textbf{857.80 ± 55.28} & 7.14 ± 0.21 \\
& Qwen2.5-7B-instruct \cite{yang2024qwen2technicalreport} & 34.30 ± 1.12 & 255.39 ± 29.22 & 6.53 ± 0.21 & 21.50 ± 1.18 & 869.08 ± 66.19 & 6.83 ± 0.26 \\
& Gemini-1.5 \cite{geminiteam2024geminifamilyhighlycapable}     & 35.47 ± 1.22 & 325.39 ± 39.61 & 5.04 ± 0.16 & 19.50 ± 1.35 & 926.05 ± 67.65 & 8.12 ± 0.31 \\
& DeepSeek-v2.5 \cite{deepseekai2024deepseekv2strongeconomicalefficient} & 34.07 ± 1.21 & 374.76 ± 33.49 & 14.50 ± 0.27 & 23.00 ± 1.32 & 945.48 ± 61.67 & 7.81 ± 0.23 \\
\midrule
\multirow{5}{*}{SearchAgent \cite{chen2024mindsearchmimickinghumanminds}}
& ChatGPT-4o \cite{yuan2024finllmsframeworkfinancialreasoning}  & 46.33 ± 1.30 & 632.11 ± 74.07 & 11.58 ± 0.06 & 27.50 ± 1.42 & 1649.86 ± 143.18 & 10.31 ± 0.16 \\
& Llama-3.1-8B-instruct \cite{touvron2023llamaopenefficientfoundation} & 43.80 ± 1.24 & 576.58 ± 42.67 & 11.61 ± 0.07 & 29.00 ± 1.31 & 1337.35 ± 122.07 & 11.42 ± 0.41 \\
& Qwen2.5-7B-instruct \cite{yang2024qwen2technicalreport}  & 41.21 ± 1.31 & 602.69 ± 56.55 & 11.51 ± 0.06 & 27.50 ± 1.74 & 1403.82 ± 132.87 & 11.81 ± 0.37 \\
& Gemini-1.5 \cite{geminiteam2024geminifamilyhighlycapable} & 42.33 ± 1.28 & 654.85 ± 53.56 & 11.58 ± 0.04 & 31.50 ± 1.14 & 1586.78 ± 141.84 & 10.72 ± 0.36 \\
& DeepSeek-v2.5 \cite{deepseekai2024deepseekv2strongeconomicalefficient} & 44.27 ± 1.26 & 697.92 ± 60.59 & 11.32 ± 0.04 & 28.50 ± 1.52 & 1635.29 ± 158.10 & 12.10 ± 0.72 \\
\midrule
\multirow{5}{*}{MindSearch \cite{chen2024mindsearchmimickinghumanminds}}
& ChatGPT-4o \cite{yuan2024finllmsframeworkfinancialreasoning} & 52.40 ± 1.33 & 3544.85 ± 473.24 & 19.09 ± 0.39 & 33.50 ± 1.12 & 4733.64 ± 356.78 & 24.12 ± 0.91 \\
& Llama-3.1-8B-instruct \cite{touvron2023llamaopenefficientfoundation} & 53.60 ± 1.28 & 3098.40 ± 277.10 & 14.82 ± 0.45 & 35.50 ± 1.41 & 4352.94 ± 337.81 & 26.14 ± 0.34 \\
& Qwen2.5-7B-instruct \cite{yang2024qwen2technicalreport} & 45.62 ± 1.26 & 3482.19 ± 298.21 & 18.12 ± 0.35 & 32.00 ± 1.17 & 4202.89 ± 316.71 & 23.42 ± 0.41 \\
& Gemini-1.5 \cite{geminiteam2024geminifamilyhighlycapable} & 51.53 ± 1.29 & 3210.20 ± 223.32 & 20.14 ± 0.33 & 36.50 ± 0.97 & 4832.15 ± 377.39 & 25.72 ± 0.58 \\
& DeepSeek-v2.5 \cite{deepseekai2024deepseekv2strongeconomicalefficient} & 49.73 ± 1.32 & 3741.20 ± 291.20 & 27.01 ± 0.58 & 33.00 ± 0.91 & 4899.24 ± 364.60 & 30.17 ± 0.93 \\
\midrule
\multirow{5}{*}{FinSearch \cite{li2024agentframeworkrealtimefinancial}}
& ChatGPT-4o  \cite{yuan2024finllmsframeworkfinancialreasoning} & 76.20 ± 1.12 & 4828.21 ± 377.59 & 16.03 ± 0.43 & 44.50 ± 2.06 & 6734.26 ± 528.45 & 19.42 ± 0.81 \\
& Llama-3.1-8B-instruct  \cite{touvron2023llamaopenefficientfoundation} & 75.53 ± 1.04 & 4572.48 ± 393.61 & 14.55 ± 0.47 & 46.50 ± 1.81 & 7060.92 ± 644.80 & 21.31 ± 0.68 \\
& Qwen2.5-7B-instruct \cite{yang2024qwen2technicalreport} & 75.40 ± 1.18 & 5482.19 ± 430.69 & 17.15 ± 0.45 & 41.00 ± 1.73 & 7120.88 ± 676.49 & 24.14 ± 0.78 \\
& Gemini-1.5 \cite{geminiteam2024geminifamilyhighlycapable} & 74.87 ± 1.08 & 5954.77 ± 376.66 & 17.74 ± 0.53 & 47.50 ± 1.47 & 7368.72 ± 695.27 & 19.71 ± 0.91 \\
& DeepSeek-v2.5 \cite{deepseekai2024deepseekv2strongeconomicalefficient} & 72.33 ± 1.15 & 5242.42 ± 458.70 & 29.31 ± 0.70 & 51.00 ± 2.42 & 7249.70 ± 706.33 & 20.73 ± 0.46 \\
\midrule
\multirow{2}{*}{SearchExpert \cite{searchexpert}}
& Llama-3.1-8B-instruct \cite{touvron2023llamaopenefficientfoundation} & 69.98 ± 0.95 & 2156.20 ± 190.83 & 12.66 ± 0.48 & 64.00 ± 2.42 & 3681.29 ± 497.56 & 15.34 ± 0.42 \\
& Qwen2.5-7B-instruct \cite{yang2024qwen2technicalreport} & 72.30 ± 0.57 & 2330.47 ± 196.89 & 13.48 ± 0.21 & 62.50 ± 1.82 & 3430.57 ± 424.59 & 14.85 ± 0.96 \\
\midrule
\textbf{R-Search}   &  DeepSeek-R1-Distill-Qwen-7B \cite{yang2024qwen2technicalreport,deepseekai2025deepseekr1incentivizingreasoningcapability}  & \textbf{78.13 ± 1.97} & 1535.55 ± 167.18 & 8.58 ± 0.49 & \textbf{73.00 ± 3.46} & 1811.34 ± 139.54 & 9.60 ± 0.68 \\
\bottomrule
\end{tabular}}
\end{table*}

\subsection{Implementation Details}\label{envdetail}
All experiments were run on a single node with 8 4090 GPUs under Python 3.10.16 and PyTorch 2.6.0 (with CUDA 12.4); we trained with a batch size per device of 1 and 4 gradient accumulation steps (effective batch size 32) using an initial learning rate of 5e-6.
We implemented R-Search using Transformers Reinforcement Learning (TRL) version 0.18.0, adapting the standard GRPO trainer to accommodate our two-stage generation process that requires environmental interaction between the \texttt{<search>} and \texttt{<result>} components. 
We leverage DeepSeek-R1-Distill-Qwen-7B \cite{deepseekai2025deepseekr1incentivizingreasoningcapability} as the LLM backbone due to its inherent reasoning capabilities that align naturally with our structured \texttt{<think>} component generation.
The automated dataset construction and reward evaluation pipeline of $\mathcal{F}_{\text{ans}}(y,a)$ leverages GPT-4.1-Mini \cite{openai2024gpt4technicalreport}.
We include three search tools for $\mathcal{T}$. 
For academic literature, we interface directly with the arXiv REST API to access recent scholarly publications. 
News content retrieval employs GNews \footnote{https://gnews.io/} as the primary endpoint with automatic fallback to Serper's GoogleSearch API \footnote{https://serper.dev/} configured in news mode when insufficient results are returned. 
General web search capabilities are provided through Serper's generic GoogleSearch API, enabling comprehensive coverage of open-web information sources.
Training efficiency optimization involves constraining each search query to retrieve the top-2 passages, reducing computational overhead during policy learning while maintaining sufficient context for effective answer synthesis. 
GRPO optimization employs a group size of $M = 4$ rollouts per query to ensure a stable advantage estimation, with KL divergence regularization coefficient $\beta$ linearly annealed from 0.1 to 0.01 over the training duration.

\subsection{Results on Search-Specific Benchmarks}

Our evaluation first employs two benchmarks specifically designed to evaluate search-augmented LLMs on temporally sensitive queries that fall outside the knowledge cutoff of pretrained LLMs.
First, FinSearchBench-24~\cite{li2024agentframeworkrealtimefinancial} comprises contemporary finance-oriented queries requiring real-time market information, presented as single-choice questions with four to five candidate answers. 
Second, SearchExpertBench-25~\cite{searchexpert} evaluates complex multi-step reasoning through more challenging multi-choice scenarios where LLMs must identify three to five correct options from typically six candidates. 
Both benchmarks source questions from recent web content to ensure evaluation of generalization capabilities beyond training distributions.
Three performance metrics are reported, namely, answer accuracy, computational efficiency measured through token consumption, and end-to-end inference latency.

As shown in Tab.~\ref{tab:benchmarks}, performance on FinSearchBench-24 demonstrates the effectiveness of R-Search. 
R-Search achieves 78.13\% accuracy, substantially outperforming the baseline without search augmentation, which achieves only 34-38\% accuracy. 
This performance advantage validates the need to search for external information for temporally sensitive financial queries. 
Notably, R-Search exceeds the domain-specific FinSearch framework (76.20\% with ChatGPT-4o), despite utilizing general-purpose rather than specialized financial search tools, while simultaneously reducing token consumption from approximately 4.8k to 1.5k tokens and decreasing inference latency from 16.0 seconds to 8.6 seconds.
The comparison with multi-agent search frameworks reveals efficiency advantages of our single-LLM approach. 
MindSearch achieves 52\% accuracy but requires 3.1k-3.7k tokens and incurs 15-27 second latency penalties due to its multi-stage architectural design.
SearchExpert with comparable model sizes (Qwen2.5-7B) reaches 72.30\% accuracy but consumes 2.3k tokens with 13.5 second latency, demonstrating that R-Search maintains superior performance while achieving substantial computational savings.
SearchExpertBench-25 evaluation further validates R-Search's scalability to complex reasoning scenarios. 
R-Search achieves 73.0\% accuracy, surpassing similarly sized implementations of SearchExpert, such as those using Llama-3.1-8B (64.0\%) and Qwen2.5-7B (62.5\%).
This performance advantage is accompanied by dramatic efficiency improvements, reducing token consumption from 3.6k to 1.8k tokens and decreasing inference latency from 15.3 seconds to 9.6 seconds. 
These results demonstrate that architectural efficiency, rather than simply increasing model parameters, provides the most effective path to improved search-augmented performance.

\subsection{Results on Q\&A Benchmarks}

\begin{table*}[htbp]
\caption{Performance comparison on single–hop and multi–hop QA benchmarks}
\label{tab:qa_benchmarks}
\centering
\resizebox{\textwidth}{!}{%
\begin{tabular}{lcccccccc}
\toprule
\multirow{2}{*}{\textbf{Method}}
& \multicolumn{3}{c}{\textbf{Single-Hop QA}}
& \multicolumn{4}{c}{\textbf{Multi-Hop QA}}
& \multirow{2}{*}{\textbf{Avg.}} \\ 
\cmidrule(lr){2-4} \cmidrule(lr){5-8}
& \textbf{NQ~\cite{NQ}} & \textbf{TriviaQA~\cite{joshi2017triviaqa}} & \textbf{PopQA~\cite{mallen2022not}}
& \textbf{HotpotQA~\cite{yang2018hotpotqa}} & \textbf{2Wiki~\cite{ho2020constructing}} & \textbf{Musique~\cite{trivedi2021musique}} & \textbf{Bamboogle~\cite{press2022measuring}} & \\ 
\midrule
Zero-Shot                   & 11.60 & 35.60 &  1.20 & 16.40 & 22.20 &  4.80 & 14.40 & 15.17 \\
CoT \cite{wei2023chainofthoughtpromptingelicitsreasoning}                             & 12.80 & 35.60 &  3.80 & 16.20 & 22.60 &  6.60 & 24.00 & 17.37 \\
RAG \cite{lewis2021retrievalaugmentedgenerationknowledgeintensivenlp}                           & 27.40 & 58.20 & 17.80 & 25.80 & 23.20 &  9.40 & 16.80 & 25.51 \\
RA-Agent  \cite{li2025search}   & 21.20 & 40.20 &  8.80 & 19.60 & 19.60 &  7.60 & 28.00 & 20.71 \\
Search-o1 \cite{li2025search}   & 19.40 & 40.60 & 11.40 & 17.00 & 27.00 &  8.60 & 30.40 & 22.06 \\
R1-base \cite{deepseekai2025deepseekr1incentivizingreasoningcapability}         & 27.60 & 47.40 & 27.40 & 21.00 & 29.20 &  9.80 & 27.78 & 27.17 \\
R1-instruct \cite{deepseekai2025deepseekr1incentivizingreasoningcapability}     & 27.00 & 45.80 & 24.20 & 21.60 & 27.80 &  8.40 & 25.00 & 25.69 \\
Search-R1-base \cite{jin2025search}      & 43.40 & 61.40 & 54.60 & 31.20 & 37.20 & 18.20 & 30.56 & 39.51 \\
Search-R1-instruct \cite{jin2025search}  & 42.40 & 63.40 & 51.60 & 32.80 & 33.20 & 17.40 & 26.39 & 38.17 \\
ZeroSearch-base \cite{sun2025zerosearch} & 42.40 & 66.40 & \textbf{60.40} & 32.00 & 34.00 & 18.00 & 33.33 & 40.93 \\
ZeroSearch-instruct \cite{sun2025zerosearch} & 43.60 & 65.20 & 48.80 & 34.60 & 35.20 & \textbf{18.40} & 27.78 & 39.08 \\
\midrule
\textbf{R-Search} & \textbf{46.20} & \textbf{69.80} & 41.60 & \textbf{35.30} & \textbf{45.20} & 15.80 & \textbf{39.20} & \textbf{41.87} \\
\bottomrule
\end{tabular}}
\end{table*}

To provide further evaluation coverage and enable direct comparison with existing single-LLM search methods, we additionally extend our assessment to include seven established Q\&A benchmarks that have been widely adopted in recent search-augmented generation research. 
Following the experimental protocols established in Search-R1~\cite{jin2025search} and ZeroSearch~\cite{sun2025zerosearch}, we evaluate R-Search on Natural Questions (NQ)~\cite{NQ}, TriviaQA~\cite{joshi2017triviaqa}, PopQA~\cite{mallen2022not}, HotpotQA~\cite{yang2018hotpotqa}, 2WikiMultihopQA~\cite{ho2020constructing}, Musique~\cite{trivedi2021musique}, and Bamboogle~\cite{press2022measuring}.  
These datasets provide evaluation scenarios that can test various aspects of factual knowledge retrieval and multi-hop reasoning. Following ZeroSearch \cite{sun2025zerosearch}, we use Exact Match (EM) as the evaluation metric, considering a prediction correct if its normalized form precisely matches any of the normalized ground-truth answers.

As shown in Table~\ref{tab:qa_benchmarks}, R-Search demonstrates effective performance across both single-hop and multi-hop QA benchmarks.
In single-hop QA tasks, R-Search achieves competitive performance on datasets such as TriviaQA, reaching 69.80\%, surpassing previous methods including ZeroSearch-base (66.40\%). 
Similarly, on the NQ dataset, R-Search attains 46.20\%, , outperforming ZeroSearch-instruct (43.60\%).
For multi-hop QA tasks, R-Search also performs effectively. 
On the 2Wiki dataset, R-Search achieves an accuracy of 45.20\%, compared to 35.20\% from ZeroSearch-instruct..
Additionally, R-Search records 39.20\% accuracy on Bamboogle, compared to 33.33\% from ZeroSearch-base. 
It also obtains improved accuracy on the HotpotQA dataset (35.30\%), compared to ZeroSearch-instruct (34.60\%).
The overall average accuracy across all seven benchmarks is 41.87\%. 
These improvements are achieved without using offline Wikipedia corpora or specialized Wikipedia-based QA training, such as those employed by Search-R1~\cite{jin2025search}, highlighting R-Search's effectiveness in handling diverse types of question-answering tasks. 

\subsection{Case Study}

\begin{figure}[htbp!]
\centering
\begin{tcolorbox}[
  colback=gray!7, colframe=black!25,
  boxrule=0.45pt, arc=2pt,
  left=3pt, right=3pt, top=3pt, bottom=3pt]
\footnotesize\ttfamily
\textcolor{orange}{\textbf{<think>}}
OK, I am now going to solve this question about the price increase of coke in \textbf{early October 2024}. First, I need to clarify the key points of the question: the time frame is \textbf{early October 2024}, the category of coke in question is \textbf{quasi-first-class metallurgical coke}, and what is its price increase. 

Next, I need to determine the appropriate search tool. Since this is a question about economic indicators and market data, the most appropriate tool should be \textbf{news search (GNews)} because it can provide the latest market dynamics and analysis reports. Web search (General) is also possible, but news search is more directly relevant. 

Then, I need to build a search query. Keywords include "\textbf{early October 2024}", "\textbf{coke quasi-first-class metallurgical coke}" and "\textbf{price increase}". These keywords are combined to accurately find relevant information... \textcolor{orange}{\textbf{</think>}}\par\medskip
\textcolor{RoyalBlue}{\textbf{<search>}}
Nodes:
A: Price fluctuation of coke (Quasi-first-grade Metallurgical Coke) in early October 2024 (News)

B: Impact of global energy market on coke prices (News)

C: International coke market dynamics in early October 2024 (News)

Edges: A -> C; B -> C \textcolor{RoyalBlue}{\textbf{</search>}} \par\medskip
\textcolor{Purple}{\textbf{<result>}}
Node A (news) Search results:

Result 1:

Title: Market price changes of important means of production in the circulation field in early October 2024 - National Bureau of Statistics

Abstract: Coke (quasi-first-grade metallurgical coke). Tons. 1768.7. 177.0. \textbf{11.1}... \textcolor{Purple}{\textbf{</result>}}\par\medskip
\textcolor{ForestGreen}{\textbf{<answer>}}
Based on the search results: 11.1\% are in line with the market supply and demand balance. The answer is: C \textcolor{ForestGreen}{\textbf{</answer>}}\par
\end{tcolorbox}

\caption{Complete R-Search inference trace demonstrating the four-component structured output framework for a temporally sensitive financial query. 
}
\label{fig:lpg-case}
\end{figure}

To demonstrate the practical effectiveness of R-Search, we present a case illustration on the complete execution flow for a temporally sensitive financial query that requires precise market information retrieval and synthesis in Fig.~\ref{fig:lpg-case}.
The selected query exemplifies the challenges that R-Search addresses: ``\textit{In early October 2024, among coal products, what is the price increase of coke (quasi-first-grade metallurgical coke)?}'', which demands specific temporal context, domain expertise, and access to recent market data that fall outside the knowledge cutoff of the LLM backbone.
The reasoning phase, encapsulated within the \texttt{<think>} component, demonstrates query decomposition and search planning. 
The LLM explicitly identifies three elements: the temporal reference point (``\textit{early October 2024}''), the specific commodity category (``\textit{quasi-first-class metallurgical coke}''), and the target metric (``\textit{price increase}'').
This structured analysis enables search plan formulation, as evidenced by the LLM's selection of news search tool over general web search based on relevance for official market statistics and economic indicators.
The search planning phase, represented by the \texttt{<search>} component, translates the reasoning insights into an executable NL-DAG structure. 
The generated plan demonstrates multi-faceted querying through three interconnected nodes: (1) direct price fluctuation inquiry, (2) global energy market impact analysis, and (3) international market dynamics assessment. 
The dependency relationships ($A\rightarrow C$; $B \rightarrow C$) illustrate how R-Search constructs hierarchical information gathering strategies that enable evidence synthesis from multiple perspectives.
The execution phase, captured in the \texttt{<result>} component, presents concrete evidence obtained through the structured search process. 
The retrieved information corresponds directly to the planned search nodes, demonstrating effective execution of the NL-DAG structure. 
The specific data point (``\textit{11.1\%}'') sourced from authoritative government statistics (``\textit{National Bureau of Statistics}'') validates the ability of R-Search to access and retrieve precise, current information from reliable institutional sources.
The synthesis phase, completed within the \texttt{<answer>} component, demonstrates evidence-based reasoning that grounds the final conclusion in the retrieved data. 
The LLM integrates the specific percentage increase with contextual market analysis to provide a definitive answer, illustrating how R-Search maintains accuracy while supporting its conclusions with verifiable evidence.

This case study validates R-Search's core value proposition: the framework successfully unifies complex reasoning, strategic multi-source planning, and accurate answer synthesis within a single coherent inference process.
The natural language representation of the search DAG proves both interpretable and executable, requiring only lightweight parsing mechanisms while maintaining the necessary for complex multi-hop information retrieval scenarios.

\subsection{Ablation Study}

\begin{table*}[htbp]
\centering
\caption{Ablation study for R-Search on multi-source search planning and ReFT}
\label{tab:ablation}
\begin{tabular}{@{}cc|ccc|ccc@{}}
\toprule
\multicolumn{2}{c|}{\textbf{Components}} & \multicolumn{3}{c|}{\textbf{FinSearchBench-24}} & \multicolumn{3}{c}{\textbf{SearchExpertBench-25}} \\
\cmidrule(lr){1-2} \cmidrule(lr){3-5} \cmidrule(lr){6-8}
\textbf{Multi-Source} & \textbf{ReFT} & \textbf{Accuracy} & \textbf{Tokens} & \textbf{Latency} & \textbf{Accuracy} & \textbf{Tokens} & \textbf{Latency} \\
\textbf{Search} & \textbf{Training} & \textbf{(\%)} & \textbf{(Count)} & \textbf{(sec)} & \textbf{(\%)} & \textbf{(Count)} & \textbf{(sec)} \\
\midrule
\ding{55} & \checkmark & 54.63 ± 4.38 & 1,316.64 ± 126.78 & 8.39 ± 0.23 & 63.00 ± 6.53 & 1,591.36 ± 97.65 & 9.67 ± 0.43 \\
\checkmark & \ding{55} & 49.28 ± 1.21 & \textbf{1,263.02 ± 58.04} & \textbf{7.95 ± 0.16} & 64.00 ± 3.97 & \textbf{1,491.21 ± 141.61} & \textbf{8.94 ± 0.40} \\
\checkmark & \checkmark & \textbf{78.13 ± 1.97} & 1,535.55 ± 167.18 & 8.58 ± 0.49 & \textbf{73.00 ± 3.46} & 1,811.34 ± 139.54 & 9.60 ± 0.68 \\
\bottomrule
\end{tabular}
\end{table*}

We conducted ablation studies to quantify the individual contributions of R-Search's two core components: multi-source search planning and ReFT in Tab.~\ref{tab:ablation}.
It reveals that both components provide substantial and complementary performance benefits. 
When multi-source search is disabled while maintaining ReFT training, accuracy on FinSearchBench-24 drops from 78.13\% to 54.63\%, representing a 23.5\% reduction, which demonstrates that access to diverse information sources is important for complex financial reasoning tasks that require synthesis of market data, regulatory information, and temporal context from multiple authoritative sources.
ReFT training is equally essential for optimal performance. 
When multi-source search is enabled but ReFT is removed, FinSearchBench-24 accuracy falls to 49.28\%, indicating a 28.85\% degradation compared to the full configuration. 
This substantial impact validates that the specialized reinforcement learning approach successfully trains the model to generate well-structured search plans while maintaining the quality of the answer synthesis through the design of the composite reward function.
The performance patterns on SearchExpertBench-25 exhibit similar trends.
Computational efficiency metrics reveal that both ablated configurations achieve modest reductions in resource consumption compared to the full framework. 
However, these marginal efficiency gains cannot justify substantial accuracy reductions that would impact practical deployment effectiveness. 
The token consumption differences remain within acceptable bounds for real-world applications, while the accuracy improvements from both components provide clear value for the end-user experience.

\section{Conclusion}\label{sec:conclusion}
This work introduces R-Search, a unified single-LLM framework for how LLM integrates search capabilities into their reasoning processes. 
By consolidating multi-step search planning and answer synthesis within a single coherent inference trace, R-Search addresses the architectural fragmentation and computational inefficiencies while overcoming the strategic limitations of current single-LLM approaches.
The major innovation lies in our structured output format that seamlessly interleaves four distinct components: reasoning traces that analyze information needs, natural-language directed acyclic graphs that encode multi-source search strategies, retrieved evidence from multiple sources, and synthesized answers grounded in factual information. 
Our reinforcement fine-tuning method introduces a composite reward function that simultaneously optimizes three dimensions to ensure that the LLM develops capabilities across all aspects of search-augmented generation rather than optimizing for accuracy alone at the expense of structural coherence.
By unifying reasoning and search within a single model, R-Search eliminates the coordination overhead and potential inconsistencies that arise from distributed multi-agent architectures. 
The natural language representation of search plans enhances interpretability, allowing practitioners to understand and debug the model's information-seeking strategies. 
Moreover, the ability to leverage multiple search sources within a single query enables more comprehensive information gathering that better mirrors human research patterns.
Nevertheless, we acknowledge certain trade-offs inherent in our design. 
The multi-source search capability, while enhancing answer quality through diverse evidence synthesis, can incur higher API costs when multiple search nodes require execution. 
Additionally, the current implementation relies on predefined search tools, limiting adaptability to novel information sources without retraining.
Looking forward, several promising future research directions emerge from this work. 
A particularly compelling avenue involves developing adaptive search strategies where the LLM dynamically assesses whether external search is necessary based on query characteristics and confidence in its parametric knowledge. 
This capability would enable more efficient resource utilization by avoiding unnecessary API calls for queries answerable from the model's training data. 
Furthermore, extending R-Search to support iterative refinement of search strategies based on initial results could enhance performance on complex queries requiring exploratory information gathering. 
%



\bibliographystyle{ACM-Reference-Format}
\bibliography{refs}

\end{document}